\documentclass[sigconf, nonacm]{acmart}

\AtBeginDocument{%
  \providecommand\BibTeX{{%
    \normalfont B\kern-0.5em{\scshape i\kern-0.25em b}\kern-0.8em\TeX}}}

\renewcommand\footnotetextcopyrightpermission[1]{} 
\usepackage{subcaption}
\usepackage{array}
\usepackage{soul}
\usepackage[textsize=footnotesize, disable]{todonotes}
\newcommand{\vect}[1]{\boldsymbol{\mathbf{#1}}}

\newcommand{\rv}[1]{\textcolor{black}{#1}}

\newcommand{\btn}[1]{\textcolor{black}{#1}} 
\begin{document}
\settopmatter{printacmref=false, printccs=false}
\setcopyright{none}
\title{Parametric Sensitivities of a Wind-driven
Baroclinic Ocean 
Using
Neural Surrogates}
\author{Yixuan Sun}
\email{yixuan.sun@anl.gov}
\orcid{}
\affiliation{%
  \institution{Mathematics and Computer Science Division, Argonne National Laboratory}
  \streetaddress{9700 S. Cass Ave.}
  \city{Lemont}
  \state{Illinois}
  \country{USA}
  \postcode{60439}
}

\author{Elizabeth Cucuzzella}
\affiliation{%
  \institution{Tufts University}
  \streetaddress{Medford}
  \city{419 Boston Ave.}
  \state{Massachusetts}
  \country{USA}}
  \postcode{02155}
\email{Elizabeth.Cucuzzella@tufts.edu}

\author{Steven Brus }
\affiliation{%
  \institution{Mathematics and Computer Science Division, Argonne National Laboratory}
  \streetaddress{9700 S. Cass Ave.}
  \city{Lemont}
  \state{Illinois}
  \country{USA}
  \postcode{60439}
}
\email{sbrus@anl.gov}

\author{Sri Hari Krishna Narayanan}
\affiliation{%
  \institution{Mathematics and Computer Science Division, Argonne National Laboratory}
  \streetaddress{9700 S. Cass Ave.}
  \city{Lemont}
  \state{Illinois}
  \country{USA}
  \postcode{60439}
}
\email{snarayan@anl.gov}

\author{Balasubramanya Nadiga}
\affiliation{%
  \institution{Los Alamos National Laboratory}
  \streetaddress{Bikini Atoll Road}
  \city{Los Alamos}
  \state{California}
  \country{USA}}
  \postcode{87545}
\email{balu@lanl.gov}

\author{Luke Van Roekel}
\affiliation{%
  \institution{Los Alamos National Laboratory}
  \streetaddress{Bikini Atoll Road}
  \city{Los Alamos}
  \state{New Mexico}
  \country{USA}}
   \postcode{87545}
\email{lvanroekel@lanl.gov}

\author{Jan H\"uckelheim}
\affiliation{%
  \institution{Mathematics and Computer Science Division, Argonne National Laboratory}
  \streetaddress{9700 S. Cass Ave.}
  \city{Lemont}
  \state{Illinois}
  \country{USA}
  \postcode{60439}
}
\email{jhuckelheim@anl.gov}

\author{Sandeep Madireddy}
\affiliation{%
  \institution{Mathematics and Computer Science Division, Argonne National Laboratory}
  \streetaddress{9700 S. Cass Ave.}
  \city{Lemont}
  \state{Illinois}
  \country{USA}
  \postcode{60439}
}
\email{smadireddy@anl.gov}

\author{Patrick Heimbach}
\affiliation{%
  \institution{University of Texas at Austin}
  \streetaddress{2515 Speedway}
  \city{Austin}
  \state{Texas}
  \country{USA}
  \postcode{78712}
}
\email{heimbach@oden.utexas.edu}
\renewcommand{\shortauthors}{Sun et al.}

\newcommand{\textanon}[1]{{\it<text removed for anonymous submission>}}
\begin{abstract}

%
    Numerical models of the ocean and ice sheets are crucial for understanding and simulating the impact of greenhouse gases on the global climate.
    Oceanic processes affect phenomena such as hurricanes, extreme precipitation, and droughts. 
    Ocean models rely on subgrid-scale parameterizations that \btn{require calibration and often} significantly affect model skill.
    When model sensitivities to parameters can be computed by using approaches such as automatic differentiation, they can be used \btn{for such calibration toward reducing}  the misfit between model output and \rv{data}. Because the \rv{SOMA model} code is challenging to differentiate,
    we have created neural network-based surrogates for estimating the sensitivity of the ocean model to model parameters. We first generated perturbed parameter ensemble data for an idealized ocean model and trained three surrogate neural network models. The neural surrogates accurately predicted the one-step forward ocean dynamics, of which we then computed the parametric sensitivity. 
\end{abstract}





\maketitle
\section{Introduction}

The oceans are important in mitigating anthropogenic climate change, primarily by absorbing significant amounts of carbon dioxide and heat. Concurrently, oceanic processes redistributing mass, heat, and salt are key drivers in phenomena such as hurricanes, extreme precipitation, and droughts.
\btn{Therefore, much} effort is devoted to modeling and understanding the behavior of the ocean under various scenarios~\cite{RN1, deyoung2004challenges, semtner1995modeling, yan2018underestimated}.
Of particular interest are the long-term changes in critical ocean circulation patterns,
which could have wide-ranging climate impacts. \rv{Understanding the stability of the circulation is critical to our ability to predict the conditions that could cause its weakening.}


We consider the Model for Prediction Across Scales MPAS-Ocean, a \rv{numerical} model designed for the simulation of the ocean system from time scales of days to millennia and spatial scales from sub-1 km to global circulations~\cite{RINGLER2013211,petersen2019,golaz2019}. \btn{Being able to represent interactions across a wide range of scales, it can directly capture mesoscale ocean activity at sufficiently high resolutions or be used to study anthropogenic climate change when used as part of a comprehensive climate or Earth system model.} 
In this context, subgrid-scale parameterization, for example, of unresolved eddies or vertical mixing, plays an important role.
We are interested in understanding the sensitivities of MPAS-Ocean's output to the model parameters. Estimating this sensitivity is time- and resource-consuming by running the model multiple times while varying each parameter in isolation. Other approaches include the \rv{use of} Gaussian process, which suffers greatly from the curse of dimensionality~\cite{hida1993gaussian, binois2022survey}, and parameter study methods~\cite{osti_1829573}.

Alternatively, adjoint models have shown great promise in uncovering the sensitivity of the model to its parameters~\cite{mcnamara_fluid_nodate,errico1992sensitivity, Stammer2005,Ferreira2005}.
They simultaneously calculate the derivatives of a single model output with respect to all model parameters.
Adjoint models can be created
manually or via automatic differentiation (AD), a technique to compute the adjoints of mathematical functions expressed as source code~\cite{Griewank2008EDP}. The reverse mode of AD, which computes adjoints, is also known as backpropagation and underpins machine learning~\cite{10.5555/3122009.3242010}. \btn{Most} popular machine learning frameworks such as PyTorch~\cite{paszke2019pytorch}, which are used to build neural networks, provide the ability to \btn{perform the reverse mode of AD and thus} compute adjoint \btn{sensitivity with little effort on the part of the end user}.
While AD tools exist for several \rv{programming} languages, an AD tool can be challenging to apply for highly parallel, pre-existing \rv{codes}, including MPAS-Ocean. 
The resulting derivative computation may exhibit poor performance and can be difficult to validate~\cite{hückelheim2023understanding}. 

Neural networks (NNs), as universal function approximators\rv{~\cite{hornik1991approximation}}, create models of physical processes from observational or simulation data and may act as surrogates for numerical models~\cite{nguyen_climax_2023, sun2023deepgraphonet, thiyagalingam2022scientific}. Once trained, NNs can infer outcomes at unseen parameter points \btn{compatible with the training data}. Recently, adjoints obtained by differentiation of the surrogate NN model \btn{have been shown to} match the accuracy of those obtained by conventional AD of the original forward simulation code in \btn{particular} settings that were considered \cite{hatfield_etal2021, chennault_etal2021}. We therefore explore how to generate an accurate NN surrogate for the idealized Simulating Ocean Mesoscale Activity (SOMA) test case within MPAS-Ocean~\cite{somaweb}. We then use our NN model to generate adjoint versions of the original model to obtain parametric sensitivities.

The contributions of this work can be summarized as follows. (1) We generated a SOMA perturbed parameter ensemble dataset for deep learning model development and benchmarking; (2) we trained neural network surrogates with large-scale distributed training, aiming to recreate the timestepping behavior of the forward (true) model; and (3) we computed and verified neural adjoints from the neural surrogates and gained insight into the model sensitivity to four parameters. 
The rest of this paper is organized as follows. Section~\ref{sec:methods} presents the SOMA simulation, problem setting, and neural network surrogate generation. Section~\ref{sec:expr} presents the experiment setup, and Section~\ref{sec:results} presents the results and discussion. Section~\ref{sec:conclusion} concludes the paper with a summary and a brief look at future work.

\section{Methods}
\label{sec:methods}
This section describes the default SOMA configuration of MPAS-Ocean, problem formulation, and neural network surrogate. 
\subsection{SOMA Configuration}
\begin{figure*}[h]
    \centering
    \includegraphics[width=.55\linewidth]{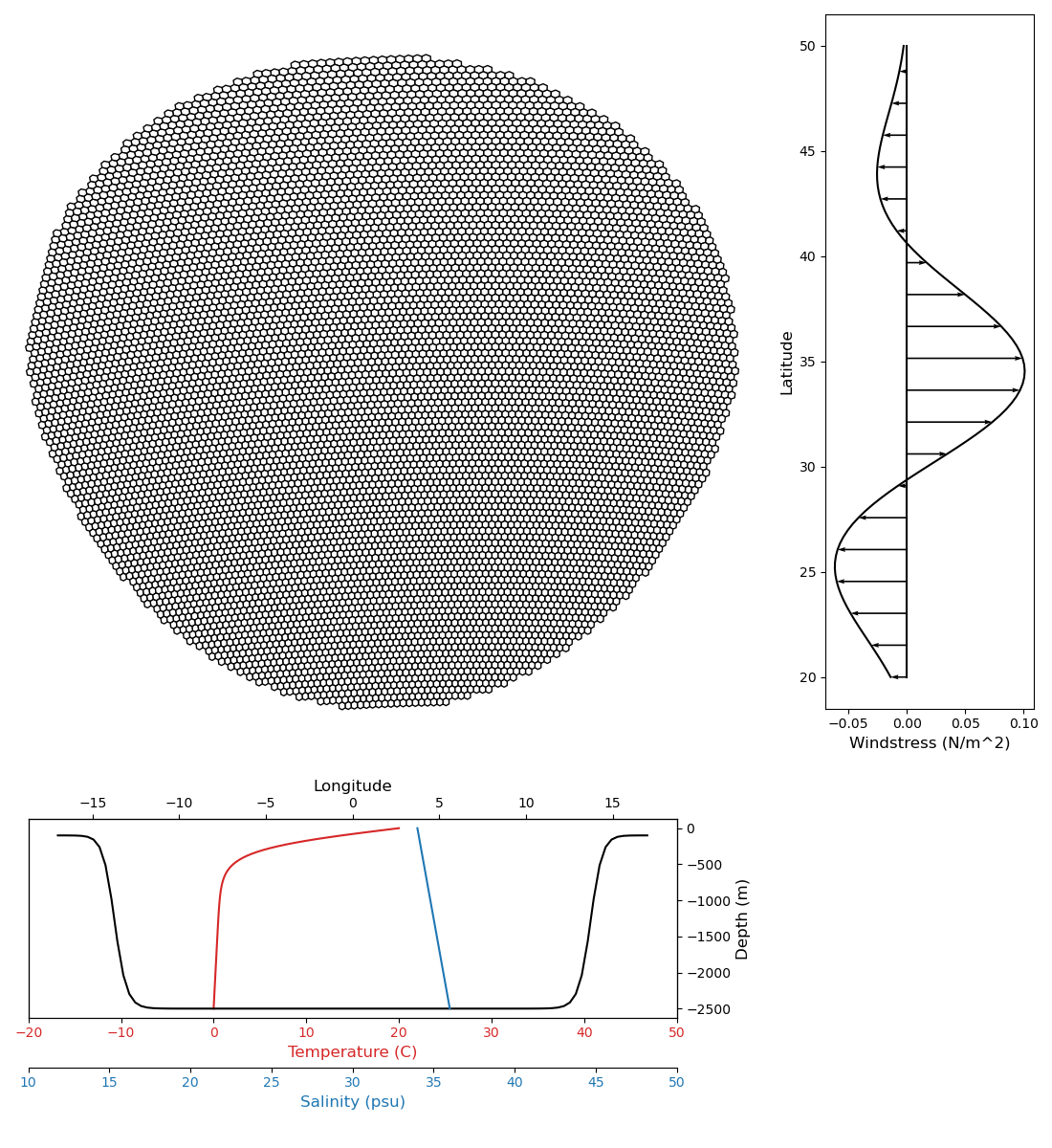}
    \caption{SOMA domain shown with the 32 km mesh. Below is the depth profile of the basin, along with the horizontally constant initial temperature and salinity profiles. To the right is the longitudinally constant (zonal-mean) imposed wind stress forcing.\label{fig:soma}}        
\end{figure*}

The SOMA configuration is designed to investigate equilibrium mesoscale eddy activity in a setting similar to how ocean climate models are deployed. SOMA simulates an idealized,
eddying, midlatitude, double-gyre ocean basin with latitudes ranging
from 21.58 to 48.58N and longitudes ranging from 16.58W to 16.58E~\cite{Wolfram2015}. The circular basin features curved coastlines
with a 150 km wide, 100 m deep continental shelf and slope (see Figure~\ref{fig:soma}). SOMA can be run at four different resolutions, where a coarser resolution is more granular: 4 km, 8 km, 16 km, and 32 km. 

We have chosen to run SOMA at a resolution of 32 km. 
At this resolution, the mesoscale eddy field is unresolved and is parametrized instead. The diagnostic output is computed from five prognostic outputs~\rv{(layer thickness, salinity, temperature, zonal velocity, and meridional velocity)} that, in turn, are influenced by four model parameters. The parameterization of mesoscale tracer transport employed in SOMA is a combination of isopycnal diffusion known as Redi parameterization~\cite{Redi1982}, \btn{with an associated parameter} $\kappa_{redi}$, and eddy-induced advection known as Gent--McWilliams parameterization~\cite{Gent1990, Gent1995}, \btn{with an associated parameter} $\kappa_{GM}$. The vertical mixing model uses the Richardson number–based parameterization~\cite{Pacanowski1981},  \btn{with an associated parameter} $\kappa_{bg}$. Bottom drag parameterization,  \btn{with an associated parameter} $C_D$, is used to extract energy supplied by wind forcing. \todo{Steven/Luke/Balu to confirm the parametrization text} The prognostic variables \btn{whose temporal evolution} we are interested in \btn{emulating} are layer thickness, salinity, meridional velocity, zonal velocity, and temperature.

\begin{figure}[h]
    \centering
    \includegraphics[width=\linewidth]{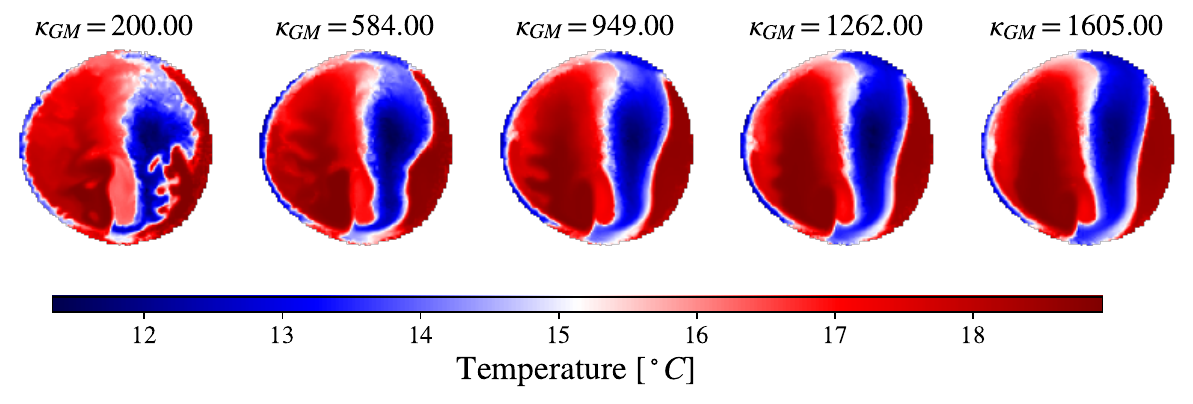}
    \caption{Variation shown in temperature with different values of Gent--McWilliams (GM) parameterization. The temperature at a depth of approximately 43 m at the end of simulations of the same initial condition is shown. \label{fig:gm_parametrization}} 
\end{figure}

The original SOMA simulation runs for constant values of the scalar parameters that are being studied. For training our NN surrogates, we are interested in varying the parameters. Figure~\ref{fig:gm_parametrization} shows the variation in ocean temperature for different values of the Gent--McWilliams parametrization. Significant variation in output for different parameter values is clearly observed. The SOMA setup is therefore suitable for creating a perturbed parameter ensemble, as described in Section~\ref{sec:datageneration}.

\todo{Steven/Luke/Balu Possibly add more text about the physical meaning of the output physical meaning of parameters. What we can learn if we know the sensitivity}


   


\subsection{Problem Setting}
\rv{This section describes the problem setup. We use $\vect{x}$ to denote the state variables, $p$ the physical model parameters, and $\theta$ the learnable parameters in the neural network.}


On a high level, SOMA starts with the initial state, $\vect{x}_0$, and the model parameter(s), $p$, and solves for the state $\vect{x}(t)$ (i.e., the prognostic variables) at time $t$. The initial value problem,
$d\vect{x}(t, p)/dt = f(\vect{x}(t))$, with $\vect{x}(0) = \vect{x}_0$, leads to 
the solution at time $t$,  
    $\vect{x}_t = \vect{x}_0 + \int_{0}^{t} f(\vect{x}(\tau, p))d \tau.$
The solution at the discrete time step $t+1$ can be expressed as 
\begin{equation}\label{eqn:one-step}
    \vect{x}(t + 1) = \vect{x}(t) + \int_{t}^{t+1}f(\vect{x}(\tau, p))d \tau.
\end{equation}
Building neural surrogates aims to model the one-step solving process in~(\ref{eqn:one-step}). We aim to use a neural network $\mathcal{N}_{\theta}$, parameterized by \rv{the neural network trainable parameters} $\theta$ to approximate the solution operator. In particular, we trained the neural network to approximate the following mapping,
\begin{equation}
    \vect{x}_{t+1} = \mathcal{N}_{\theta}(\vect{x}_t, p),
\end{equation}
where $\vect{x}$ is the state variable vector, $t$ represents the current time, and $p$
is the parameter that impacts the trajectories of state variables. \rv{The dataset $\mathcal{D}$, obtained from each ensemble with varying $p$, described in Section~\ref{sec:datageneration}, contains the input-output pairs $\{[\vect{x}_t^{(k)},p^{(k)}], \vect{x}^{(k)}_{t+1}\}^N_{k=1}$. Now, the learning objective is to minimize the empirical loss function},
\begin{equation}
    \mathcal{L}({\theta; \mathcal{D}}) = \frac{1}{N}\sum_{k=1}^N \Vert \vect{x}_{t+1}^{(k)} - \mathcal{N}_{\theta}(\vect{x}_t^{(k)}, p^{(k)})\Vert^2_2,
\end{equation}
where $N$ is the number of data points in the training set. We expect the trained neural network to accurately predict the state variables one step forward. Furthermore, with an accurate neural surrogate for the forward process, we are interested in understanding the adjoint sensitivity of the model with respect to the parameter $p$, $\partial \mathcal{N}_{\theta} / \partial p$, 
\todo{Aren't we interested in the gradient $\nabla_{\vect{p}}$?} \todo{YS: the output of NNs is not a scalar, so I think using Jacobian is more appropriate.}
\todo{SHK: We do not have a scalar-valued cost function at the moment, and we do not have a spatially varying input field. So a gradient/reverse mode is not what we need to compute, although that is what we are doing.}
which can be easily calculated through backpropagation, \rv{once a skillful, trained neural network is available}. 
\rv{Because the SOMA simulation is not readily differentiable using AD, we verify the calculation of neural adjoints by performing a dot-product test, described in Section~\ref{sec:dottest}, as the first step toward matching the model true adjoint $\partial \mathcal{M}/ \partial p$, when a differentiable physical model $\mathcal{M}$ is available.}


\subsection{Neural Network Surrogates}

\begin{table*}[]
    \centering
    \caption{The normalized root mean square error~(\%) of the predicted prognostic variables from U-Net, ResNet, and FNO trained for $\kappa_{GM}$. }
    \begin{tabular}{lccccc}
    \toprule
         &  Layer Thickness & Salinity & Temperature & Zonal Velocity & Meridional Velocity\\
    \midrule
        U-Net & 0.8961  & 0.6288 & 2.190 & 0.8679  & 0.5518\\
        ResNet& 4.273 &  3.978 &  3.889 &  1.624 &  1.914\\
        FNO & 0.2944 & 0.1706 & 0.2176 & 0.3195 & 0.2059\\
    \bottomrule
    \end{tabular}
    
    \label{tab:prelim}
\end{table*}

\begin{figure*}
    \centering
    \includegraphics[width=.9\linewidth]{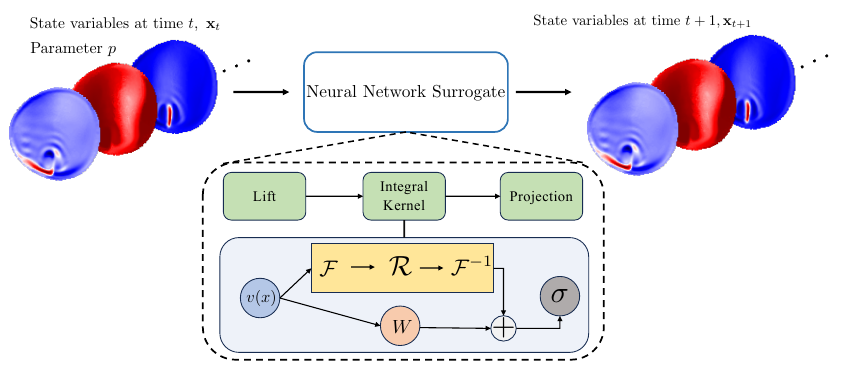}
    \caption{Fourier neural operator surrogate schematic for one-step forward forecasting of prognostic variables. The input state variable fields and parameters are in 3D representations. The lift block performs pointwise projection to raise the dimensions of the input. A Fourier layer performs fast Fourier transform~(FFT) of the input and conducts convolution operations in the frequency domain. The high-frequency modes from the convolution are truncated. At the end of the Fourier layer, an inverse FFT transforms the data onto its original physical domain. Another layer of pointwise projection then aligns with the dimension of the solution function.}
    \label{fig:NN}
\end{figure*}

We trained three types of NNs that are commonly used in learning dynamical systems~\cite{nguyen2023climax, bonev2023spherical}---U-Net~\cite{ronneberger2015u}, ResNet~\cite{he2016deep}, and FNO~\cite{li2021fourier}---to learn the forward dynamics of the simulation. The values of normalized root mean square error (NRMSE) of the predicted prognostic variables from the three NNs trained for $\kappa_{GM}$, listed in Table~\ref{tab:prelim}, show that the Fourier neural operator (FNO) greatly outperforms~(with lower NRMSEs) the other two NNs. We therefore primarily adopted FNO as the surrogate model to train on data from the dynamics of all varied parameters. FNO aims to learn the operator between infinite-dimensional function spaces and \rv{can handle high-dimensional data efficiently in the spectral space.} FNO has had many successes in modeling physical systems~\cite{rashid2022learning, pathak2022fourcastnet, li2022fourier}. \rv{In particular, FourCastNet~\cite{pathak2022fourcastnet}, powered by FNO backbones, achieved outstanding performance in short- to mid-range global weather forecasting.} The key component of FNO is the kernel integral operator implemented as a convolution in the Fourier domain. By compositing such kernel integral operators along with nonlinear lift and projection operations, FNO effectively approximates such an operator and can generalize beyond the data resolution in the training set. Figure~\ref{fig:NN} shows the data flow with FNO in our problem, where the input consists of 3D representations of the state variables and the external parameter, $\vect{x}$, $p$. The output is the same state variables at the next time step. \rv{We zero-padded the region outside the circular domain, mitigating the edge effect and avoiding artifacts from the periodic boundary assumption for the fast Fourier transform.}

In the Fourier layer, the integral kernel performs convolution in the frequency domain, capturing global dependencies. In particular, high-frequency modes are filtered out to preserve the main structure of the solution, increasing the generalization ability. Because of the high-dimensional nature of our data~(three-dimensional in space with multiple variables), we use a variant of FNO, called TFNO~\cite{kossaifi2023multigrid}, which incorporates Tucker decomposition for more efficient learning. \rv{Tucker decomposition essentially does a principal component analysis for higher-order tensors~\cite{kolda2009tensor}}, reducing the complexity of the trainable parameters in the FNO backbones, resulting in better training efficiency and generalization~\cite{kossaifi2023multigrid}.

\section{Experiments}
This section describes the data generation and postprocessing, neural network training, and evaluation metrics for the trained surrogate models.
\label{sec:expr}

\subsection{Perturbed Parameter Ensemble Data Generation and Transformation}
\label{sec:datageneration}
 \rv{We derived from the literature a set of reasonable ranges for perturbing the parameters.}
 Table~\ref{tab:parameterrange} lists the parameters and maximum and minimum values for uniform sampling.
 \rv{We independently perturbed each parameter in the list and performed simulations to form one ensemble per parameter. For each ensemble, we trained a neural network surrogate.}
\rv{For each parameter}, using uniformly sampled values in the range, we created 100 forward runs.  In each run, we executed the simulation from the same initial condition with different parameter configurations for three years and saved the \textit{daily} \rv{snapshots} of the last year. \rv{At the 32 km resolution, there are 8,521 hexagonal cells on the grid, each with 60 vertical levels. Each month resulted in over 15 million ($8521 \times 60 \times 30$) cell values for each spatially and temporally varying output variable in the data set.} Each run of the simulation was done 
with 128 cores via MPI. The problem setup and simulation code can be found on \href{https://github.com/anl-ecucuzzella/SOMAForwardCode}{GitHub}.

\begin{table}[h]
    \centering
    \caption{Range of Perturbed Parameter Values.}
    \begin{tabular}{cccc}
    \toprule 
    \textbf{SOMA Parameter} & \textbf{Symbol} &\textbf{Minimum} & \textbf{Maximum}\\
    \midrule
      GM\_constant\_kappa & $\kappa_{GM}$ & 200.0 & 2000.0 \\
      Redi\_constant\_kappa & $\kappa_{redi}$ & 0.0 & 3000.0\\
      cvmix\_background\_diff & $\kappa_{bg}$ & 0.0 & 1e-4\\ 
      implicit\_bottom\_drag & $C_{D}$& 1e-4 & 1e-2\\
    \bottomrule
    \end{tabular}
   
    \label{tab:parameterrange}
\end{table}

\rv{The generated data with its original fidelity and representation posed challenges for training FNO-based models.}
Therefore, the data generated for the mesh grid was converted to a standard latitude and longitude grid through spatial interpolation, and the values were populated to standard array entries. As a result, we obtained the data represented with regular grids and stored as arrays, each instance of shape (6, 60, 100, 100). The first dimension contains the five prognostic variables and one model parameter, and the last three dimensions represent the spatial axes of the domain. We postprocessed the \textit{first month} of saved data for neural surrogate training and evaluation purposes. In total, we have 3,000 \rv{data instances~(30 time steps $\times$ 100 runs)} for \rv{the ensemble per perturbed parameter.}

\subsection{Neural Surrogate Training}
We implemented the data loaders, data transformations, and TFNOs in PyTorch~\cite{paszke2019pytorch}. The data loaders streamed the input-output pairs in batches, alleviating memory usage. Based on \rv{independent} runs of SOMA, the data was randomly split into training, validation, and testing sets with a ratio of 0.6, 0.2, and 0.2. \rv{That is, the training set contained 60 forward runs, and 20 forward runs for both validation and testing sets. }  Our objective was to predict all five prognostic variables from a single model simultaneously. \rv{The state variables lie in dramatically different ranges. The range difference can cause the learning to be biased toward certain variables. Therefore, we normalized the data to ensure the resulting values had similar ranges. In particular, we used the minimum and maximum variable values from the simulation instances to perform the normalization with a target range of [0, 1]. }
With the high dimensionality of the data and complexity of the neural network, we distributed the training and data loader onto ten computing nodes equipped with 4 Nvidia A100 GPUs each. We trained all four models with a batch size of 10, $L_2$ norm as loss function, and Adam optimizer~\cite{kingma2017adam} for \rv{over 2,000 epochs with observed convergence.} Then, \rv{based on the validation loss}, we selected the best model snapshot with the lowest validation loss for further evaluation with the testing set. The implementation and training scripts are available on \href{https://github.com/iamyixuan/DeepAdjoint}{GitHub}.

\subsection{Metrics}

We use three metrics to evaluate the predictions of the model based on ground truth: coefficient of determination~($R^2$),  NRMSE, and anomaly correlation coefficient~(ACC). $R^2$ is defined as $R^2 = 1 - SS_{res}/SS_{tot}$, where $SS_{res} = \sum(y - \hat{y})^2$; $SS_{tot} = \sum(y - \bar{y})^2$; and $y$, $\hat{y}$, and $\bar{y}$ are true values, predicted value, and the average of true values, respectively.
It describes the proportion of variance in the target that the model can explain. A number closer to 1 is associated with a better model. NRMSE is defined as $NRMSE = \frac{1}{N}\sqrt{\sum(y - \hat{y})^2} / (y_{max} - y_{min})$. The NRMSE accounts for the average percentage deviation of the predicted values from the true values. It is normalized by using the true data range to compare different models and predictions for different state variables fairly. The third metric, ACC, implies the similarity between two state variable fields, including the pattern-contrast variations. It is defined as $ACC = \sum(y - \bar{y})(\hat{y} - \bar{y})/ \sqrt{\sum(y - \bar{y})^2(\hat{y} - \bar{y})^2}$~\cite{von1997analysis}.
ACC ranges from -1 to 1, with 1 indicating a perfect positive correlation between the anomalies from predicted and true values. All three metrics are robust to state variable types and scales.

\section{Results and Discussion}
This section discusses the accuracy of our trained NNs and the computation of adjoints. 
\label{sec:results}
\subsection{Forward Prediction}

\begin{table*}
    \centering
    \caption{Single-step forward prediction results (TFNO): Varying $\kappa_{GM}$, $\kappa_{redi}$, $\kappa_{bg}$, and $C_D$.} \label{tab:model_performance}
    \centering
    \begin{tabular}{lcccccc}
    \toprule

        ~ & \multicolumn{3}{c}{$\kappa_{GM}$} & \multicolumn{3}{c}{$\kappa_{redi}$}\\
        \cmidrule{2-4}\cmidrule{5-7}
        ~  & $R^2$ & NRMSE~(\%) & ACC & $R^2$ & NRMSE~(\%) & ACC\\
        \midrule
        Layer Thickness     & 0.9999 & 0.2944 & 0.9999  & 0.9987 & 0.6119 & 0.9993 \\ 
        Salinity            & 0.9999 & 0.1706 & 1.000   & 0.9993 & 0.4992 & 0.9997 \\ 
        Temperature         & 0.9999 & 0.2176 & 1.000   & 0.9921 & 0.1028 & 0.9960 \\  
        Zonal Vel.          & 0.9611 & 0.3195 & 0.9806  & 0.9314 & 0.2007 & 0.9806 \\ 
        Meridional Vel.     & 0.9928 & 0.2059 & 0.9969  & 0.9346 & 0.1903 & 0.9969 \\ 
        \midrule
        ~ & \multicolumn{3}{c}{$\kappa_{bg}$} & \multicolumn{3}{c}{$C_D$}\\
        \cmidrule{2-4}\cmidrule{5-7}
        ~  & $R^2$ & NRMSE~(\%) & ACC & $R^2$ & NRMSE~(\%) & ACC\\
        \midrule
        Layer Thickness     & 0.9997 & 0.3463 & 0.9998 & 0.9997 & 0.4050 & 0.9990 \\ 
        Salinity            & 0.9998 & 0.3439 & 0.9999 & 0.9999 & 0.2281 & 1.000 \\ 
        Temperature         & 0.9998 & 0.3649 & 0.9999 & 0.9998 & 0.3652 & 0.9999 \\ 
        Zonal Vel.          & 0.9789 & 0.4271 & 0.9894 & 0.9546 & 0.4780 & 0.9799 \\ 
        Meridional Vel.     & 0.9838 & 0.4020 & 0.9919 & 0.9828 & 0.3749 & 0.9914 \\  
        \bottomrule
    \end{tabular}
\end{table*}

Table~\ref{tab:model_performance} shows the performance metrics of the TFNOs for each output prognostic variable ~(layer thickness, salinity, temperature, zonal velocity, and meridional velocity) with varying parameters. The trained models, overall, accurately predict the prognostic variables one step in the future. The model performance on zonal and meridional velocities among the four trained networks is less accurate than other variables. In particular, the network taking varying $\kappa_{bg}$ is the least performant in predicting zonal and meridional velocities. One possible reason is that zonal and meridional velocities have high contrast profiles in small regions, whereas other variables lack \rv{sharp changes in values}. Since we trained the network to minimize the total loss for all variables, it was more challenging for the network to capture high-frequency features present in zonal and meridional velocity profiles. 


\begin{figure*}
    \centering
    \includegraphics[width=.9\linewidth]{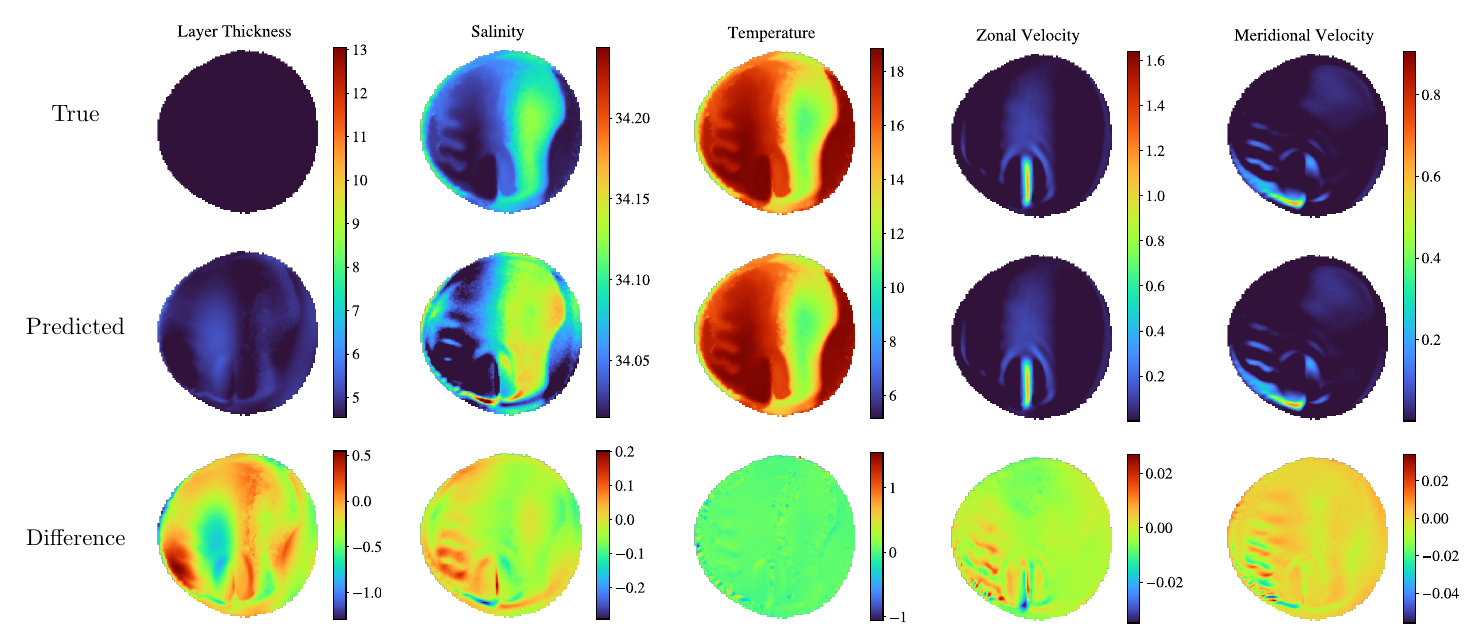}
    \caption{One-step forward predictions from the trained TFNO for all five prognostic variables at a depth of approximately 43 m and at timestep 15 (days). 
    Units are (from left to right column): m, g/kg,  $^{\circ}$C, m/s, m/s.}
    \label{fig:onestep}
\end{figure*}

\begin{figure*}[thbp]
    \centering
    \begin{subfigure}{.33\textwidth}
    \centering
    \includegraphics[width=\linewidth]{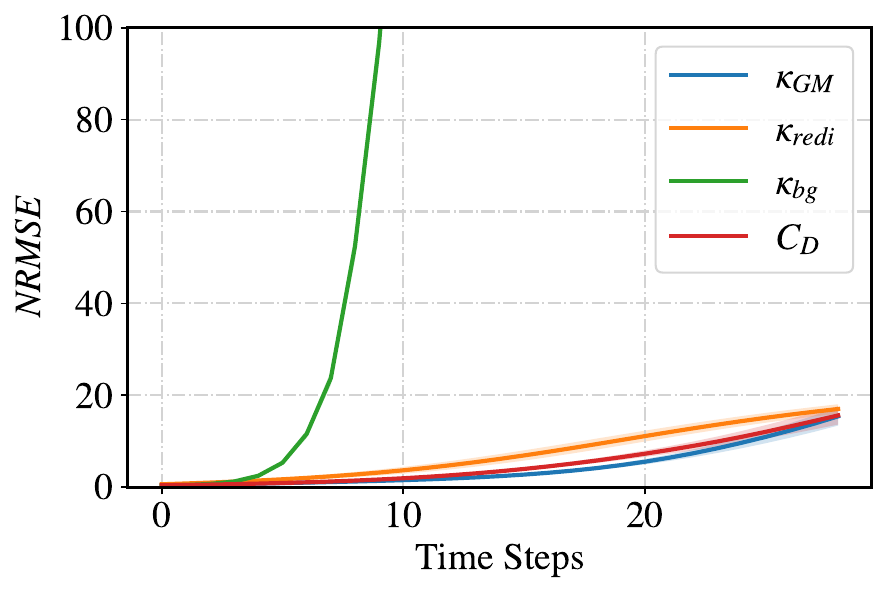}
    \caption{Layer Thickness}
    \end{subfigure}
    \begin{subfigure}{.33\textwidth}
    \centering
    \includegraphics[width=\linewidth]{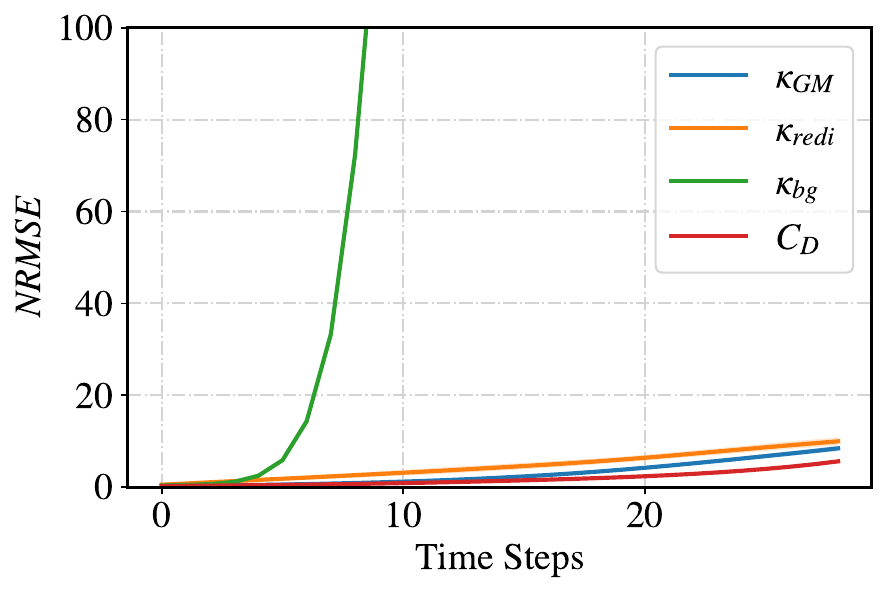}
    \caption{Salinity}
    \end{subfigure}
    \begin{subfigure}{.33\textwidth}
    \centering
    \includegraphics[width=\linewidth]{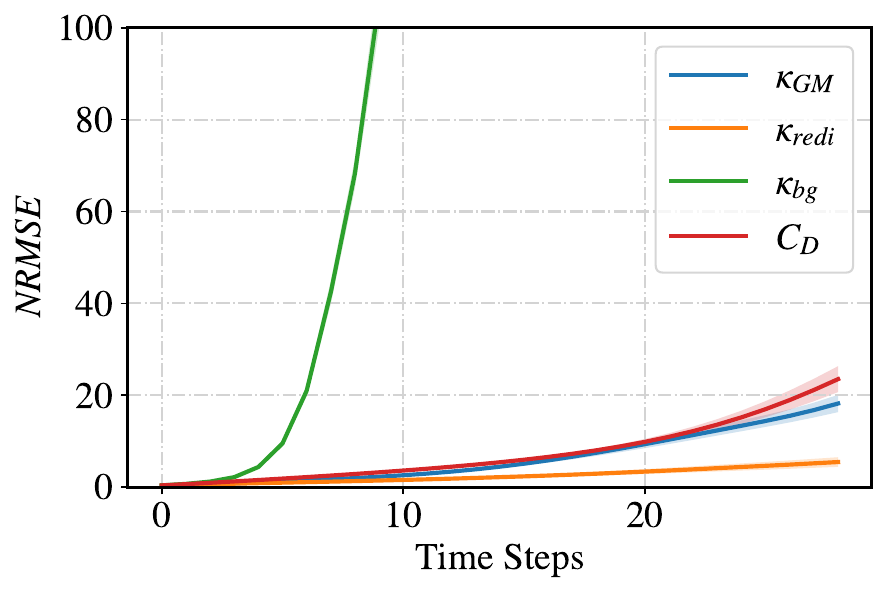}
    \caption{Temperature}
    \end{subfigure}
    
    \begin{subfigure}{.33\textwidth}
    \centering
    \includegraphics[width=\linewidth]{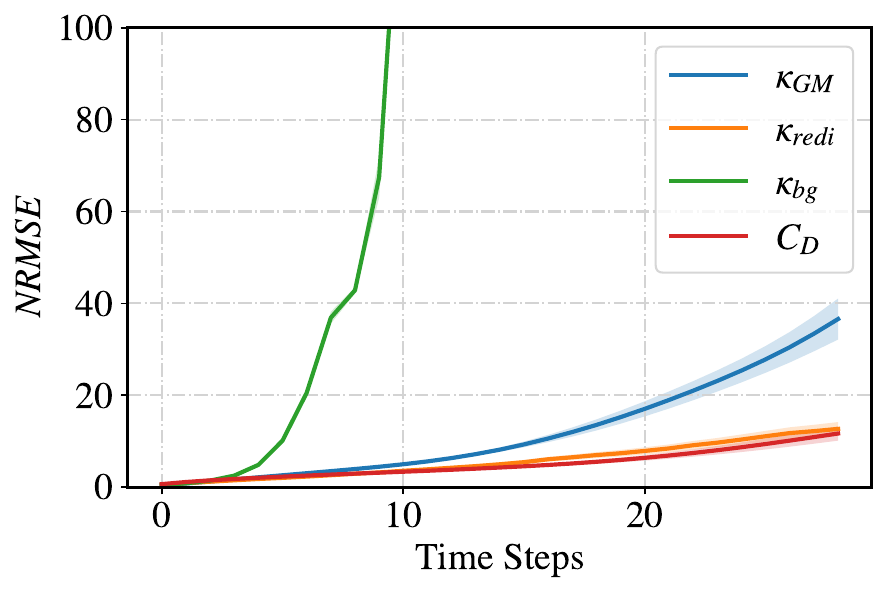}
    \caption{Zonal Velocity}
    \end{subfigure}
    \begin{subfigure}{.33\textwidth}
    \centering
    \includegraphics[width=\linewidth]{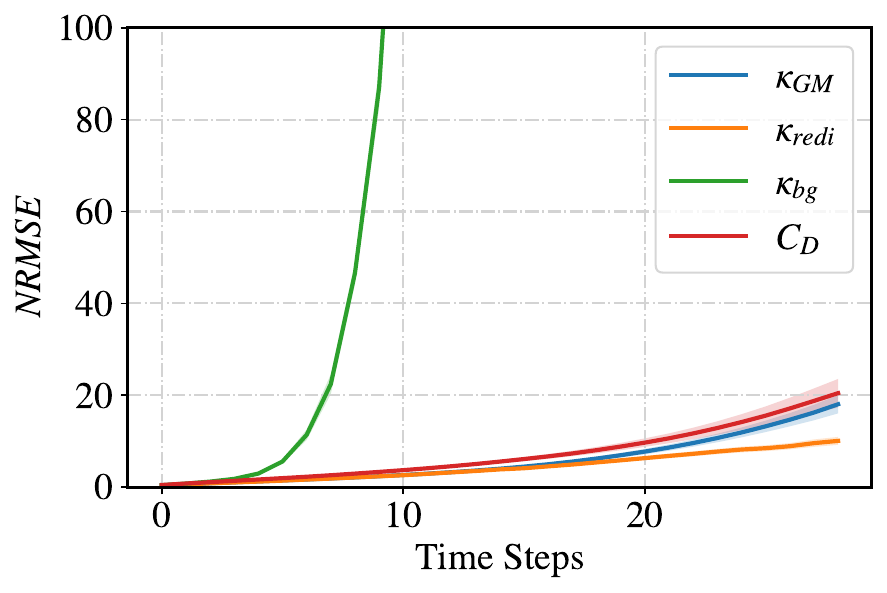}
    \caption{Meridional Velocity}
    \end{subfigure}
    \caption{Rollout performance, reported in NRMSE, of the trained surrogates for the prognostic variables. The neural surrogates for varying $\kappa_{GM}$, $\kappa_{redi}$, and $C_D$ make accurate predictions in an autoregressive way up to 20 steps (days) until the error accumulation starts to rapidly increase. Meanwhile, the performance of neural surrogate for varying $\kappa_{bg}$ quickly degrades in the first few steps.}
    \label{fig:rollout}
\end{figure*}

Figure~\ref{fig:onestep} shows the true, predicted, and absolute error for one-step forward values for the prognostic variables of the trained TFNO for the parameter, $\kappa_{GM}$. The predicted profiles follow the same range of variable values and closely resemble the actual distribution, with a slight loss of details. Moreover, we are interested in the model performance for more extended horizon rollout, where the trained networks predict prognostic variables in an autoregressive way. Figure~\ref{fig:rollout} presents the four neural surrogates' rollout performance on the testing set. The NRMSE indicates accurate rollout from three of the four surrogates more than 10 steps~(days), and the model taking varying $\kappa_{redi}$ has the least accumulated error for temperature and meridional velocity profiles.  The model taking $\kappa_{bg}$ exhibits the fastest error accumulation compared with other models, holding up a relatively accurate rollout of only about 5 steps. This could be due to the trained neural surrogate not fully capturing the possibly more intricate and nonlinear interactions between the prognostic variables and $\kappa_{bg}$, resulting in a small error~(comparable to other models) in single-step forward prediction but large error accumulation in the multistep rollout.

\subsection{Neural Adjoints }

We are focused on understanding how sensitive the prognostic variables are to the parameter using the neural surrogates. 
Using the accurate neural surrogate
makes it possible to determine the adjoint sensitivity through backpropagation.  We have computed the sensitivity of the neural surrogates to the parameters of the physical model by calculating the Jacobian of the surrogate's output w.r.t. the parameter. The output field is of shape $\vect{x}_{t+1} \in \mathbb{R}^{n \times 60 \times 100 \times 100}$, where $n$ is the number of state variables~(five in our case). The input \rv{physical} model parameters, which do not vary spatially, have the shape of $p \in \mathbb{R}^{m}$, where $m=1$ is the number of model parameters in our case. 
The Jacobian, $\frac{\partial \mathcal{N}_{\theta}}{\partial (\vect{x_t}, p)} $ ,of the the trained NNs, $\mathcal{N}_{\theta}:~ \mathbb{R}^{(n+m) \times 60 \times 100 \times 100} \mapsto \mathbb{R}^{n \times 60 \times 100 \times 100} $, is computationally expensive to compute given its size. 
However, to obtain the adjoint sensitivity to the parameter, $\frac{\partial \mathcal{N}_{\theta}}{\partial p} $, by differentiating the NNs involves differentiating w.r.t the state variables as well, because the parameters and state are stored in a single tensor.
Therefore, we calculated the Jacobians 
at various randomly selected locations in the output domain
and obtained
the gradient with respect to $p$, per location, per state variable. It leads to calculating the gradient of a function, $\mathcal{N}^*_{\theta}:~\mathbb{R}^{(n+m) \times 60 \times 100 \times 100} \mapsto  \mathbb{R}^{1}$, which is straightforward and efficient using backpropagation. Then, we extracted the portion related to $p$. Because of its spatial uniformity, for a selected location and prognostic variable, we had $J_{{loc.}} \in \mathbb{R}^{1 \times m}$ as the measure of the sensitivity.
These Jacobians were used to show the sensitivity of output state variables to the \rv{physical} model parameters, summarized in Table~\ref{tab:adjoints}. The values were the averaged gradient of the state variables to the parameters $\kappa_{GM}$, $\kappa_{redi}$, $\kappa_{bg}$, and $C_D$ over 5 randomly selected horizontal locations with a fixed vertical level in the domain and were extended for over a month in the testing set. The results suggest, for the trained neural surrogate, that the outputs are least affected by $\kappa_{GM}$ since its sensitivity values are an order of magnitude less than those of the other parameters. For the neural surrogates with varying $\kappa_{GM}$ and $\kappa_{redi}$,  we observe higher sensitivities for temperature, zonal, and meridional velocity values than for layer thickness and salinity. 
\todo{For salinity, not a good range was picked. In the colorscale, I'd try a range [29, 36]}
Zonal and meridional velocity sensitivities to $\kappa_{bg}$ are the highest among the prognostic variables, while temperature takes the lowest value. For the neural surrogate taking $C_D$, the sensitivities of all prognostic variables are at a similar level.\todo{Steven, Luke, Balue: Can you comment on what this means for SOMA}

\begin{table*}[]
    \centering
    \caption{Average adjoint sensitivities computed for the five prognostic variables and the four parameters. The average was calculated across five randomly selected horizontal locations at a fixed vertical level (depth of approximately 43 m) within the domain and extended over a period of 30 days.}
    \begin{tabular}{lccccc }
    \toprule
         & Layer Thickness & Salinity & Temperature & Zonal Velocity & Meridional Velocity\\
    \midrule
        ${\partial \mathcal{N}}/{\partial \kappa_{GM}}$ & 1.642e-09 & 2.353e-09 & 3.379e-09 & 2.615e-09 & 3.014e-09\\ 
        ${\partial \mathcal{N}}/{\partial \kappa_{redi}}$ & 1.521e-08 & 2.319e-08 & 6.857e-08 & 9.578e-08 & 8.248e-08\\ 
        ${\partial \mathcal{N}}/{\partial \kappa_{bg}}$ & 3.920e-08 & 5.233e-08 & 2.321e-08 & 6.345e-08 & 7.051e-08\\ 
        ${\partial \mathcal{N}}/{\partial C_D}$ & 1.886e-08 & 2.367e-08 & 1.995e-08 & 1.848e-08 & 2.413e-08\\ 
    \bottomrule
    \end{tabular}
    
    \label{tab:adjoints}
\end{table*}

\subsubsection{Adjoint Dot-Product Tests}\label{sec:dottest}
We aim to examine the correctness of the calculated neural adjoints. 
The Jacobian of the neural surrogate at a specific spatial location has the form $J_{loc} = \frac{\partial \mathcal{N}_{\theta}}{\partial h_N} \frac{\partial h_N}{\partial h_{N-1}} \dots \frac{\partial h_0}{\partial p} \in \mathbb{R}^{n \times m}$, where $h$s are the outputs of the hidden layers in the NN. Surrogate adjoints are easily accessible by reverse-mode differentiation of trained neural networks. \rv{Given the complexity of the neural surrogate and data representations, we decided to implement a test that compares the neural adjoints with the Jacobian obtained by direct differentiation~(forward differentiation). To this end, we formulated a dot-product test that allows us to efficiently verify the consistency between the two differentiation modes}, shown in Equation~\ref{eqn:dotp-test}. 
In particular, with random vectors $\vect{v} \in \mathbb{R}^{m}$ 
and $\vect{w} \in \mathbb{R}^{n}$, we tested whether the equality sign holds. Instead of explicitly calculating the Jacobian matrices, which can \rv{require a large amount of memory and be computationally prohibitive}, we focused on computing the vector-Jacobian product~(VJP), associated with adjoint calculation, and the Jacobian-vector product~(JVP), associated with direct differentiation. We then generated 100 random vector $\vect{w}, \vect{v}$ pairs and compared the results from the left-hand side~(LHS) and right-hand side~(RHS) of Equation~\ref{eqn:dotp-test}.
\begin{equation}\label{eqn:dotp-test}
    \begin{aligned}
        &w^{\top}\cdot\Big(\underbrace{\frac{\partial \mathcal{N}_{\theta}}{\partial p}v}_{\text{JVP}}\Big) \stackrel{?}{=} \Big(\underbrace{\underbrace{w^{\top}\frac{\partial \mathcal{N}_{\theta}}{\partial h_N}} \frac{\partial h_N}{\partial h_{N-1}} \dots \frac{\partial h_0}{\partial p}}_{\text{VJP}}\Big)\cdot v, \\
    \end{aligned}
\end{equation}

\begin{figure*}[]
    \centering
        \includegraphics[width=\linewidth]{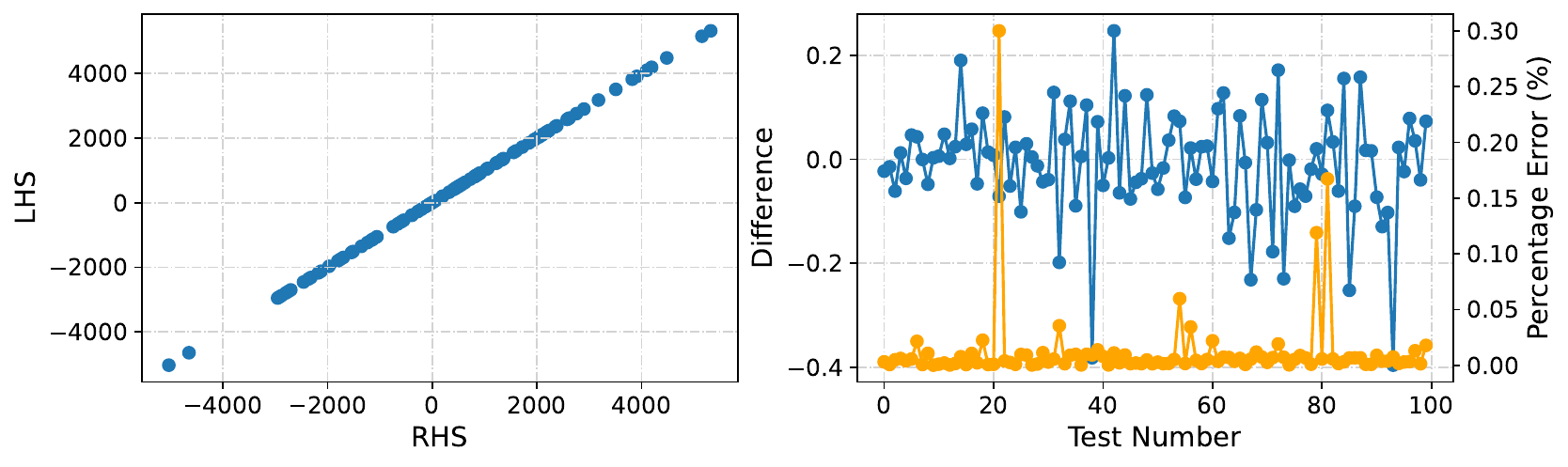}
    \caption{Dot-product test results of the trained neural surrogate. In Equation~\ref{eqn:dotp-test}, the right-hand side calculates the neural adjoints via reverse-mode differentiation. The left-hand side uses forward differentiation to calculate the Jacobian-vector product. $\vect{w}$ and $\vect{v}$ are random vectors. The difference between the two sides and the corresponding percentage error suggest a good match between the neural adjoint and direct differentiation.}
    \label{fig:dottest}
\end{figure*}
Figure~\ref{fig:dottest} shows the scatter plot of the LHS~(direct differentiation) and RHS~(adjoint calculation). The data points form a visually straight line, suggesting the calculated values are close or equal on both sides. Furthermore, the difference curve~(in blue) and percentage error curve~(in orange) show great alignment of the neural adjoints and direct differentiation, verifying the neural adjoint sensitivity based on the accurate surrogate. 

\section{Conclusion}
\label{sec:conclusion}
Based on perturbed parameters from SOMA runs, the neural surrogates we developed accurately predict the prognostic variables in one step forward compared with the original SOMA model. Although trained for one-step forward predictions, three of four neural surrogates can produce accurate rollouts of up to 10 timesteps. Additionally, we have successfully computed the neural adjoints from the trained models and obtained initial insights into the sensitivities. The results show that zonal and meridional velocity profiles are the most sensitive to $\kappa_{GM}$, $\kappa_{redi}$, and $\kappa_{bg}$. At the same time, the temperature is more sensitive to $\kappa_{GM}$ and $\kappa_{redi}$. 

Our future work includes improving adjoint-aware training by incorporating known physics and investigating the feasibility of applying our methodology to the MPAS-O code, specifically in a configuration simulating the AMOC. The governing equations of MPAS-O are well established. Utilizing known physics in the training of neural surrogates helps improve accuracy, reduce the requirement of large data size, and regularize learning for better generalization~\cite{li_physics-informed_2022, karniadakis2021physics, raissi_physics_2017, liu2022predicting}. 


\begin{acks}
We gratefully acknowledge the computing resources provided on Bebop, a high-performance computing cluster operated by LCRC at Argonne National Laboratory.
This research used resources from the NERSC, a U.S. Department of Energy Office of Science User Facility located at LBNL.
Material based upon work supported by the U.S. Department of Energy, Office of Science, Office of Advanced Scientific Computing Research and Office of BER, Scientific Discovery through Advanced Computing (SciDAC) program, under Contract DE-AC02-06CH11357. We are grateful to the Sustainable Horizons Institute's Sustainable Research Pathways workforce development program.
\end{acks}

\bibliographystyle{ACM-Reference-Format}
\bibliography{refs}


\begin{thebibliography}{46}


\ifx \showCODEN    \undefined \def \showCODEN     #1{\unskip}     \fi
\ifx \showDOI      \undefined \def \showDOI       #1{#1}\fi
\ifx \showISBNx    \undefined \def \showISBNx     #1{\unskip}     \fi
\ifx \showISBNxiii \undefined \def \showISBNxiii  #1{\unskip}     \fi
\ifx \showISSN     \undefined \def \showISSN      #1{\unskip}     \fi
\ifx \showLCCN     \undefined \def \showLCCN      #1{\unskip}     \fi
\ifx \shownote     \undefined \def \shownote      #1{#1}          \fi
\ifx \showarticletitle \undefined \def \showarticletitle #1{#1}   \fi
\ifx \showURL      \undefined \def \showURL       {\relax}        \fi
\providecommand\bibfield[2]{#2}
\providecommand\bibinfo[2]{#2}
\providecommand\natexlab[1]{#1}
\providecommand\showeprint[2][]{arXiv:#2}

\bibitem[som({[n.\,d.]})]%
        {somaweb}
 \bibinfo{year}{[n.\,d.]}\natexlab{}.
\newblock \bibinfo{title}{ocean/soma test group}.
\newblock \bibinfo{howpublished}{\url{https://mpas-dev.github.io/compass/latest/users_guide/ocean/test_groups/soma.html}}.
\newblock
\newblock
\shownote{Accessed: 2023-11-29}.


\bibitem[Adams et~al\mbox{.}(2021)]%
        {osti_1829573}
\bibfield{author}{\bibinfo{person}{Brian~M. Adams}, \bibinfo{person}{William~J. Bohnhoff}, \bibinfo{person}{Keith~R. Dalbey}, \bibinfo{person}{Mohamed~S. Ebeida}, \bibinfo{person}{John~P. Eddy}, \bibinfo{person}{Michael~S. Eldred}, \bibinfo{person}{Russell~W. Hooper}, \bibinfo{person}{Patricia~D. Hough}, \bibinfo{person}{Kenneth~T. Hu}, \bibinfo{person}{John~D. Jakeman}, \bibinfo{person}{Mohammad Khalil}, \bibinfo{person}{Kathryn~A. Maupin}, \bibinfo{person}{Jason~A. Monschke}, \bibinfo{person}{Elliott~M. Ridgway}, \bibinfo{person}{Ahmad~A. Rushdi}, \bibinfo{person}{Daniel~Thomas Seidl}, \bibinfo{person}{John~Adam Stephens}, {and} \bibinfo{person}{Justin~G. Winokur}.} \bibinfo{year}{2021}\natexlab{}.
\newblock \showarticletitle{Dakota, A Multilevel Parallel Object-Oriented Framework for Design Optimization, Parameter Estimation, Uncertainty Quantification, and Sensitivity Analysis: Version 6.15 User's Manual}.
\newblock  (\bibinfo{date}{11} \bibinfo{year}{2021}).
\newblock
\urldef\tempurl%
\url{https://doi.org/10.2172/1829573}
\showDOI{\tempurl}


\bibitem[Baydin et~al\mbox{.}(2017)]%
        {10.5555/3122009.3242010}
\bibfield{author}{\bibinfo{person}{At\i{}l\i{}m~G\"{u}nes Baydin}, \bibinfo{person}{Barak~A. Pearlmutter}, \bibinfo{person}{Alexey~Andreyevich Radul}, {and} \bibinfo{person}{Jeffrey~Mark Siskind}.} \bibinfo{year}{2017}\natexlab{}.
\newblock \showarticletitle{Automatic Differentiation in Machine Learning: A Survey}.
\newblock \bibinfo{journal}{\emph{J. Mach. Learn. Res.}} \bibinfo{volume}{18}, \bibinfo{number}{1} (\bibinfo{date}{Jan.} \bibinfo{year}{2017}), \bibinfo{pages}{5595–5637}.
\newblock
\showISSN{1532-4435}


\bibitem[Binois and Wycoff(2022)]%
        {binois2022survey}
\bibfield{author}{\bibinfo{person}{Mickael Binois} {and} \bibinfo{person}{Nathan Wycoff}.} \bibinfo{year}{2022}\natexlab{}.
\newblock \showarticletitle{A survey on high-dimensional {Gaussian} process modeling with application to {Bayesian} optimization}.
\newblock \bibinfo{journal}{\emph{ACM Transactions on Evolutionary Learning and Optimization}} \bibinfo{volume}{2}, \bibinfo{number}{2} (\bibinfo{year}{2022}), \bibinfo{pages}{1--26}.
\newblock


\bibitem[Bonev et~al\mbox{.}(2023)]%
        {bonev2023spherical}
\bibfield{author}{\bibinfo{person}{Boris Bonev}, \bibinfo{person}{Thorsten Kurth}, \bibinfo{person}{Christian Hundt}, \bibinfo{person}{Jaideep Pathak}, \bibinfo{person}{Maximilian Baust}, \bibinfo{person}{Karthik Kashinath}, {and} \bibinfo{person}{Anima Anandkumar}.} \bibinfo{year}{2023}\natexlab{}.
\newblock \showarticletitle{Spherical {Fourier} Neural Operators: Learning Stable Dynamics on the Sphere}.
\newblock \bibinfo{journal}{\emph{arXiv preprint arXiv:2306.03838}} (\bibinfo{year}{2023}).
\newblock


\bibitem[Chennault et~al\mbox{.}(2021)]%
        {chennault_etal2021}
\bibfield{author}{\bibinfo{person}{Austin Chennault}, \bibinfo{person}{Andrey~A. Popov}, \bibinfo{person}{Amit~N. Subrahmanya}, \bibinfo{person}{Rachel Cooper}, \bibinfo{person}{Anuj Karpatne}, {and} \bibinfo{person}{Adrian Sandu}.} \bibinfo{year}{2021}\natexlab{}.
\newblock \showarticletitle{Adjoint-Matching Neural Network Surrogates for Fast {4D-Var} Data Assimilation}.
\newblock \bibinfo{journal}{\emph{CoRR}}  \bibinfo{volume}{abs/2111.08626} (\bibinfo{year}{2021}).
\newblock
\urldef\tempurl%
\url{https://arxiv.org/abs/2111.08626}
\showURL{%
\tempurl}


\bibitem[deYoung et~al\mbox{.}(2004)]%
        {deyoung2004challenges}
\bibfield{author}{\bibinfo{person}{Brad deYoung}, \bibinfo{person}{Mike Heath}, \bibinfo{person}{Francisco Werner}, \bibinfo{person}{Fei Chai}, \bibinfo{person}{Bernard Megrey}, {and} \bibinfo{person}{Patrick Monfray}.} \bibinfo{year}{2004}\natexlab{}.
\newblock \showarticletitle{Challenges of Modeling Ocean Basin Ecosystems}.
\newblock \bibinfo{journal}{\emph{Science}} \bibinfo{volume}{304}, \bibinfo{number}{5676} (\bibinfo{year}{2004}), \bibinfo{pages}{1463--1466}.
\newblock
\urldef\tempurl%
\url{https://doi.org/10.1126/science.1094858}
\showDOI{\tempurl}


\bibitem[Errico and Vukicevic(1992)]%
        {errico1992sensitivity}
\bibfield{author}{\bibinfo{person}{Ronald~M Errico} {and} \bibinfo{person}{Tomislava Vukicevic}.} \bibinfo{year}{1992}\natexlab{}.
\newblock \showarticletitle{Sensitivity analysis using an adjoint of the {PSU-NCAR} mesoseale model}.
\newblock \bibinfo{journal}{\emph{Monthly Weather Review}} \bibinfo{volume}{120}, \bibinfo{number}{8} (\bibinfo{year}{1992}), \bibinfo{pages}{1644--1660}.
\newblock
\urldef\tempurl%
\url{https://doi.org/10.1175/1520-0493(1992)120<1644:SAUAAO>2.0.CO;2}
\showDOI{\tempurl}


\bibitem[Ferreira et~al\mbox{.}(2005)]%
        {Ferreira2005}
\bibfield{author}{\bibinfo{person}{David Ferreira}, \bibinfo{person}{John Marshall}, {and} \bibinfo{person}{Patrick Heimbach}.} \bibinfo{year}{2005}\natexlab{}.
\newblock \showarticletitle{Estimating Eddy Stresses by Fitting Dynamics to Observations Using a Residual-Mean Ocean Circulation Model and Its Adjoint}.
\newblock \bibinfo{journal}{\emph{Journal of Physical Oceanography}} \bibinfo{volume}{35}, \bibinfo{number}{10} (\bibinfo{year}{2005}), \bibinfo{pages}{1891 -- 1910}.
\newblock
\urldef\tempurl%
\url{https://doi.org/10.1175/JPO2785.1}
\showDOI{\tempurl}


\bibitem[Gent and Mcwilliams(1990)]%
        {Gent1990}
\bibfield{author}{\bibinfo{person}{Peter~R. Gent} {and} \bibinfo{person}{James~C. Mcwilliams}.} \bibinfo{year}{1990}\natexlab{}.
\newblock \showarticletitle{Isopycnal Mixing in Ocean Circulation Models}.
\newblock \bibinfo{journal}{\emph{Journal of Physical Oceanography}} \bibinfo{volume}{20}, \bibinfo{number}{1} (\bibinfo{year}{1990}), \bibinfo{pages}{150--155}.
\newblock
\urldef\tempurl%
\url{https://doi.org/10.1175/1520-0485(1990)020<0150:IMIOCM>2.0.CO;2}
\showDOI{\tempurl}


\bibitem[Gent et~al\mbox{.}(1995)]%
        {Gent1995}
\bibfield{author}{\bibinfo{person}{Peter~R. Gent}, \bibinfo{person}{Jurgen Willebrand}, \bibinfo{person}{Trevor~J. McDougall}, {and} \bibinfo{person}{James~C. McWilliams}.} \bibinfo{year}{1995}\natexlab{}.
\newblock \showarticletitle{Parameterizing Eddy-Induced Tracer Transports in Ocean Circulation Models}.
\newblock \bibinfo{journal}{\emph{Journal of Physical Oceanography}} \bibinfo{volume}{25}, \bibinfo{number}{4} (\bibinfo{year}{1995}), \bibinfo{pages}{463 -- 474}.
\newblock
\urldef\tempurl%
\url{https://doi.org/10.1175/1520-0485(1995)025<0463:PEITTI>2.0.CO;2}
\showDOI{\tempurl}


\bibitem[Golaz et~al\mbox{.}(2019)]%
        {golaz2019}
\bibfield{author}{\bibinfo{person}{Jean-Christophe Golaz}, \bibinfo{person}{Peter~M. Caldwell}, \bibinfo{person}{Luke~P. Van~Roekel}, \bibinfo{person}{Mark~R. Petersen}, \bibinfo{person}{Qi Tang}, \bibinfo{person}{Jonathan~D. Wolfe}, \bibinfo{person}{Guta Abeshu}, \bibinfo{person}{Valentine Anantharaj}, \bibinfo{person}{Xylar~S. Asay-Davis}, \bibinfo{person}{David~C. Bader}, \bibinfo{person}{Sterling~A. Baldwin}, \bibinfo{person}{Gautam Bisht}, \bibinfo{person}{Peter~A. Bogenschutz}, \bibinfo{person}{Marcia Branstetter}, \bibinfo{person}{Michael~A. Brunke}, \bibinfo{person}{Steven~R. Brus}, \bibinfo{person}{Susannah~M. Burrows}, \bibinfo{person}{Philip~J. Cameron-Smith}, \bibinfo{person}{Aaron~S. Donahue}, \bibinfo{person}{Michael Deakin}, \bibinfo{person}{Richard~C. Easter}, \bibinfo{person}{Katherine~J. Evans}, \bibinfo{person}{Yan Feng}, \bibinfo{person}{Mark Flanner}, \bibinfo{person}{James~G. Foucar}, \bibinfo{person}{Jeremy~G. Fyke}, \bibinfo{person}{Brian~M. Griffin}, \bibinfo{person}{Cécile Hannay},
  \bibinfo{person}{Bryce~E. Harrop}, \bibinfo{person}{Mattthew~J. Hoffman}, \bibinfo{person}{Elizabeth~C. Hunke}, \bibinfo{person}{Robert~L. Jacob}, \bibinfo{person}{Douglas~W. Jacobsen}, \bibinfo{person}{Nicole Jeffery}, \bibinfo{person}{Philip~W. Jones}, \bibinfo{person}{Noel~D. Keen}, \bibinfo{person}{Stephen~A. Klein}, \bibinfo{person}{Vincent~E. Larson}, \bibinfo{person}{L.~Ruby Leung}, \bibinfo{person}{Hong-Yi Li}, \bibinfo{person}{Wuyin Lin}, \bibinfo{person}{William~H. Lipscomb}, \bibinfo{person}{Po-Lun Ma}, \bibinfo{person}{Salil Mahajan}, \bibinfo{person}{Mathew~E. Maltrud}, \bibinfo{person}{Azamat Mametjanov}, \bibinfo{person}{Julie~L. McClean}, \bibinfo{person}{Renata~B. McCoy}, \bibinfo{person}{Richard~B. Neale}, \bibinfo{person}{Stephen~F. Price}, \bibinfo{person}{Yun Qian}, \bibinfo{person}{Philip~J. Rasch}, \bibinfo{person}{J.~E.~Jack Reeves~Eyre}, \bibinfo{person}{William~J. Riley}, \bibinfo{person}{Todd~D. Ringler}, \bibinfo{person}{Andrew~F. Roberts}, \bibinfo{person}{Erika~L. Roesler},
  \bibinfo{person}{Andrew~G. Salinger}, \bibinfo{person}{Zeshawn Shaheen}, \bibinfo{person}{Xiaoying Shi}, \bibinfo{person}{Balwinder Singh}, \bibinfo{person}{Jinyun Tang}, \bibinfo{person}{Mark~A. Taylor}, \bibinfo{person}{Peter~E. Thornton}, \bibinfo{person}{Adrian~K. Turner}, \bibinfo{person}{Milena Veneziani}, \bibinfo{person}{Hui Wan}, \bibinfo{person}{Hailong Wang}, \bibinfo{person}{Shanlin Wang}, \bibinfo{person}{Dean~N. Williams}, \bibinfo{person}{Phillip~J. Wolfram}, \bibinfo{person}{Patrick~H. Worley}, \bibinfo{person}{Shaocheng Xie}, \bibinfo{person}{Yang Yang}, \bibinfo{person}{Jin-Ho Yoon}, \bibinfo{person}{Mark~D. Zelinka}, \bibinfo{person}{Charles~S. Zender}, \bibinfo{person}{Xubin Zeng}, \bibinfo{person}{Chengzhu Zhang}, \bibinfo{person}{Kai Zhang}, \bibinfo{person}{Yuying Zhang}, \bibinfo{person}{Xue Zheng}, \bibinfo{person}{Tian Zhou}, {and} \bibinfo{person}{Qing Zhu}.} \bibinfo{year}{2019}\natexlab{}.
\newblock \showarticletitle{The {DOE E3SM} Coupled Model Version 1: Overview and Evaluation at Standard Resolution}.
\newblock \bibinfo{journal}{\emph{Journal of Advances in Modeling Earth Systems}} \bibinfo{volume}{11}, \bibinfo{number}{7} (\bibinfo{year}{2019}), \bibinfo{pages}{2089--2129}.
\newblock
\urldef\tempurl%
\url{https://doi.org/10.1029/2018MS001603}
\showDOI{\tempurl}


\bibitem[Griewank and Walther(2008)]%
        {Griewank2008EDP}
\bibfield{author}{\bibinfo{person}{Andreas Griewank} {and} \bibinfo{person}{Andrea Walther}.} \bibinfo{year}{2008}\natexlab{}.
\newblock \bibinfo{booktitle}{\emph{Evaluating Derivatives: {P}rinciples and Techniques of Algorithmic Differentiation} (\bibinfo{edition}{2nd} ed.)}.
\newblock Number 105 in \bibinfo{series}{Other Titles in Applied Mathematics}. \bibinfo{publisher}{SIAM}, \bibinfo{address}{Philadelphia, PA}.
\newblock
\showISBNx{978--0--898716--59--7}
\urldef\tempurl%
\url{http://bookstore.siam.org/ot105/}
\showURL{%
\tempurl}


\bibitem[Hatfield et~al\mbox{.}(2021)]%
        {hatfield_etal2021}
\bibfield{author}{\bibinfo{person}{Sam Hatfield}, \bibinfo{person}{Matthew Chantry}, \bibinfo{person}{Peter Dueben}, \bibinfo{person}{Philippe Lopez}, \bibinfo{person}{Alan Geer}, {and} \bibinfo{person}{Tim Palmer}.} \bibinfo{year}{2021}\natexlab{}.
\newblock \showarticletitle{Building Tangent-Linear and Adjoint Models for Data Assimilation With Neural Networks}.
\newblock \bibinfo{journal}{\emph{Journal of Advances in Modeling Earth Systems}} \bibinfo{volume}{13}, \bibinfo{number}{9} (\bibinfo{year}{2021}), \bibinfo{pages}{e2021MS002521}.
\newblock
\urldef\tempurl%
\url{https://doi.org/10.1029/2021MS002521}
\showDOI{\tempurl}


\bibitem[He et~al\mbox{.}(2016)]%
        {he2016deep}
\bibfield{author}{\bibinfo{person}{Kaiming He}, \bibinfo{person}{Xiangyu Zhang}, \bibinfo{person}{Shaoqing Ren}, {and} \bibinfo{person}{Jian Sun}.} \bibinfo{year}{2016}\natexlab{}.
\newblock \showarticletitle{Deep residual learning for image recognition}. In \bibinfo{booktitle}{\emph{Proceedings of the IEEE conference on computer vision and pattern recognition}}. \bibinfo{pages}{770--778}.
\newblock


\bibitem[Hida and Hitsuda(1993)]%
        {hida1993gaussian}
\bibfield{author}{\bibinfo{person}{Takeyuki Hida} {and} \bibinfo{person}{Masuyuki Hitsuda}.} \bibinfo{year}{1993}\natexlab{}.
\newblock \bibinfo{booktitle}{\emph{Gaussian processes}}. Vol.~\bibinfo{volume}{120}.
\newblock \bibinfo{publisher}{American Mathematical Soc.}
\newblock


\bibitem[Hornik(1991)]%
        {hornik1991approximation}
\bibfield{author}{\bibinfo{person}{Kurt Hornik}.} \bibinfo{year}{1991}\natexlab{}.
\newblock \showarticletitle{Approximation capabilities of multilayer feedforward networks}.
\newblock \bibinfo{journal}{\emph{Neural networks}} \bibinfo{volume}{4}, \bibinfo{number}{2} (\bibinfo{year}{1991}), \bibinfo{pages}{251--257}.
\newblock


\bibitem[Hückelheim et~al\mbox{.}(2023)]%
        {hückelheim2023understanding}
\bibfield{author}{\bibinfo{person}{Jan Hückelheim}, \bibinfo{person}{Harshitha Menon}, \bibinfo{person}{William Moses}, \bibinfo{person}{Bruce Christianson}, \bibinfo{person}{Paul Hovland}, {and} \bibinfo{person}{Laurent Hascoët}.} \bibinfo{year}{2023}\natexlab{}.
\newblock \bibinfo{title}{Understanding Automatic Differentiation Pitfalls}.
\newblock
\newblock
\showeprint[arxiv]{2305.07546}~[math.NA]


\bibitem[IPCC(2021)]%
        {RN1}
\bibfield{author}{\bibinfo{person}{IPCC}.} \bibinfo{year}{2021}\natexlab{}.
\newblock \bibinfo{booktitle}{\emph{Climate Change 2021: The Physical Science Basis. Contribution of Working Group {I} to the Sixth Assessment Report of the Intergovernmental Panel on Climate Change}}. Vol.~\bibinfo{volume}{In Press}.
\newblock \bibinfo{publisher}{Cambridge University Press}, \bibinfo{address}{Cambridge, United Kingdom and New York, NY, USA}.
\newblock
\urldef\tempurl%
\url{https://doi.org/10.1017/9781009157896}
\showDOI{\tempurl}


\bibitem[Karniadakis et~al\mbox{.}(2021)]%
        {karniadakis2021physics}
\bibfield{author}{\bibinfo{person}{George~Em Karniadakis}, \bibinfo{person}{Ioannis~G Kevrekidis}, \bibinfo{person}{Lu Lu}, \bibinfo{person}{Paris Perdikaris}, \bibinfo{person}{Sifan Wang}, {and} \bibinfo{person}{Liu Yang}.} \bibinfo{year}{2021}\natexlab{}.
\newblock \showarticletitle{Physics-informed machine learning}.
\newblock \bibinfo{journal}{\emph{Nature Reviews Physics}} \bibinfo{volume}{3}, \bibinfo{number}{6} (\bibinfo{year}{2021}), \bibinfo{pages}{422--440}.
\newblock
\urldef\tempurl%
\url{https://doi.org/10.1038/s42254-021-00314-5}
\showDOI{\tempurl}


\bibitem[Kingma and Ba(2017)]%
        {kingma2017adam}
\bibfield{author}{\bibinfo{person}{Diederik~P. Kingma} {and} \bibinfo{person}{Jimmy Ba}.} \bibinfo{year}{2017}\natexlab{}.
\newblock \bibinfo{title}{Adam: A Method for Stochastic Optimization}.
\newblock
\newblock
\showeprint[arxiv]{1412.6980}~[cs.LG]


\bibitem[Kolda and Bader(2009)]%
        {kolda2009tensor}
\bibfield{author}{\bibinfo{person}{Tamara~G Kolda} {and} \bibinfo{person}{Brett~W Bader}.} \bibinfo{year}{2009}\natexlab{}.
\newblock \showarticletitle{Tensor decompositions and applications}.
\newblock \bibinfo{journal}{\emph{SIAM Rev.}} \bibinfo{volume}{51}, \bibinfo{number}{3} (\bibinfo{year}{2009}), \bibinfo{pages}{455--500}.
\newblock


\bibitem[Kossaifi et~al\mbox{.}(2023)]%
        {kossaifi2023multigrid}
\bibfield{author}{\bibinfo{person}{Jean Kossaifi}, \bibinfo{person}{Nikola~Borislavov Kovachki}, \bibinfo{person}{Kamyar Azizzadenesheli}, {and} \bibinfo{person}{Anima Anandkumar}.} \bibinfo{year}{2023}\natexlab{}.
\newblock \bibinfo{title}{Multi-Grid Tensorized {Fourier} Neural Operator for High Resolution {PDE}s}.
\newblock
\newblock
\urldef\tempurl%
\url{https://openreview.net/forum?id=po-oqRst4Xm}
\showURL{%
\tempurl}


\bibitem[Li et~al\mbox{.}(2021)]%
        {li2021fourier}
\bibfield{author}{\bibinfo{person}{Zongyi Li}, \bibinfo{person}{Nikola Kovachki}, \bibinfo{person}{Kamyar Azizzadenesheli}, \bibinfo{person}{Burigede Liu}, \bibinfo{person}{Kaushik Bhattacharya}, \bibinfo{person}{Andrew Stuart}, {and} \bibinfo{person}{Anima Anandkumar}.} \bibinfo{year}{2021}\natexlab{}.
\newblock \bibinfo{title}{Fourier Neural Operator for Parametric Partial Differential Equations}.
\newblock
\newblock
\showeprint[arxiv]{2010.08895}~[cs.LG]


\bibitem[Li et~al\mbox{.}(2022)]%
        {li2022fourier}
\bibfield{author}{\bibinfo{person}{Zhijie Li}, \bibinfo{person}{Wenhui Peng}, \bibinfo{person}{Zelong Yuan}, {and} \bibinfo{person}{Jianchun Wang}.} \bibinfo{year}{2022}\natexlab{}.
\newblock \showarticletitle{Fourier neural operator approach to large eddy simulation of three-dimensional turbulence}.
\newblock \bibinfo{journal}{\emph{Theoretical and Applied Mechanics Letters}} \bibinfo{volume}{12}, \bibinfo{number}{6} (\bibinfo{year}{2022}), \bibinfo{pages}{100389}.
\newblock


\bibitem[Li et~al\mbox{.}({[n.\,d.]})]%
        {li_physics-informed_2022}
\bibfield{author}{\bibinfo{person}{Zongyi Li}, \bibinfo{person}{Hongkai Zheng}, \bibinfo{person}{Nikola Kovachki}, \bibinfo{person}{David Jin}, \bibinfo{person}{Haoxuan Chen}, \bibinfo{person}{Burigede Liu}, \bibinfo{person}{Kamyar Azizzadenesheli}, {and} \bibinfo{person}{Anima Anandkumar}.} \bibinfo{year}{[n.\,d.]}\natexlab{}.
\newblock \bibinfo{title}{Physics-Informed Neural Operator for Learning Partial Differential Equations}.
\newblock
\newblock
\urldef\tempurl%
\url{https://doi.org/10.48550/arXiv.2111.03794}
\showDOI{\tempurl}


\bibitem[Liu et~al\mbox{.}(2022)]%
        {liu2022predicting}
\bibfield{author}{\bibinfo{person}{Xin-Yang Liu}, \bibinfo{person}{Hao Sun}, \bibinfo{person}{Min Zhu}, \bibinfo{person}{Lu Lu}, {and} \bibinfo{person}{Jian-Xun Wang}.} \bibinfo{year}{2022}\natexlab{}.
\newblock \bibinfo{title}{Predicting parametric spatiotemporal dynamics by multi-resolution PDE structure-preserved deep learning}.
\newblock
\newblock
\showeprint[arxiv]{2205.03990}~[cs.LG]


\bibitem[Lyu et~al\mbox{.}(2018)]%
        {Stammer2005}
\bibfield{author}{\bibinfo{person}{Guokun Lyu}, \bibinfo{person}{Armin Köhl}, \bibinfo{person}{Ion Matei}, {and} \bibinfo{person}{Detlef Stammer}.} \bibinfo{year}{2018}\natexlab{}.
\newblock \showarticletitle{Adjoint-Based Climate Model Tuning: Application to the {Planet Simulator}}.
\newblock \bibinfo{journal}{\emph{Journal of Advances in Modeling Earth Systems}} \bibinfo{volume}{10}, \bibinfo{number}{1} (\bibinfo{year}{2018}), \bibinfo{pages}{207--222}.
\newblock
\urldef\tempurl%
\url{https://doi.org/10.1002/2017MS001194}
\showDOI{\tempurl}
\showeprint{https://agupubs.onlinelibrary.wiley.com/doi/pdf/10.1002/2017MS001194}


\bibitem[McNamara et~al\mbox{.}(2004)]%
        {mcnamara_fluid_nodate}
\bibfield{author}{\bibinfo{person}{Antoine McNamara}, \bibinfo{person}{Adrien Treuille}, \bibinfo{person}{Zoran Popovi\'{c}}, {and} \bibinfo{person}{Jos Stam}.} \bibinfo{year}{2004}\natexlab{}.
\newblock \showarticletitle{Fluid Control Using the Adjoint Method}.
\newblock \bibinfo{journal}{\emph{ACM Trans. Graph.}} \bibinfo{volume}{23}, \bibinfo{number}{3} (\bibinfo{date}{aug} \bibinfo{year}{2004}), \bibinfo{pages}{449–--456}.
\newblock
\showISSN{0730-0301}
\urldef\tempurl%
\url{https://doi.org/10.1145/1015706.1015744}
\showDOI{\tempurl}


\bibitem[Nguyen et~al\mbox{.}({[n.\,d.]})]%
        {nguyen_climax_2023}
\bibfield{author}{\bibinfo{person}{Tung Nguyen}, \bibinfo{person}{Johannes Brandstetter}, \bibinfo{person}{Ashish Kapoor}, \bibinfo{person}{Jayesh~K. Gupta}, {and} \bibinfo{person}{Aditya Grover}.} \bibinfo{year}{[n.\,d.]}\natexlab{}.
\newblock \bibinfo{title}{{ClimaX}: A foundation model for weather and climate}.
\newblock
\newblock
\showeprint[arxiv]{2301.10343 [cs]}
\urldef\tempurl%
\url{http://arxiv.org/abs/2301.10343}
\showURL{%
\tempurl}
\newblock
\shownote{version: 1}.


\bibitem[Nguyen et~al\mbox{.}(2023)]%
        {nguyen2023climax}
\bibfield{author}{\bibinfo{person}{Tung Nguyen}, \bibinfo{person}{Johannes Brandstetter}, \bibinfo{person}{Ashish Kapoor}, \bibinfo{person}{Jayesh~K Gupta}, {and} \bibinfo{person}{Aditya Grover}.} \bibinfo{year}{2023}\natexlab{}.
\newblock \showarticletitle{ClimaX: A foundation model for weather and climate}.
\newblock \bibinfo{journal}{\emph{arXiv preprint arXiv:2301.10343}} (\bibinfo{year}{2023}).
\newblock


\bibitem[Pacanowski and Philander(1981)]%
        {Pacanowski1981}
\bibfield{author}{\bibinfo{person}{R.~C. Pacanowski} {and} \bibinfo{person}{S.~G.~H. Philander}.} \bibinfo{year}{1981}\natexlab{}.
\newblock \showarticletitle{Parameterization of Vertical Mixing in Numerical Models of Tropical Oceans}.
\newblock \bibinfo{journal}{\emph{Journal of Physical Oceanography}} \bibinfo{volume}{11}, \bibinfo{number}{11} (\bibinfo{year}{1981}), \bibinfo{pages}{1443--1451}.
\newblock
\urldef\tempurl%
\url{https://doi.org/10.1175/1520-0485(1981)011<1443:POVMIN>2.0.CO;2}
\showDOI{\tempurl}


\bibitem[Paszke et~al\mbox{.}(2019)]%
        {paszke2019pytorch}
\bibfield{author}{\bibinfo{person}{Adam Paszke}, \bibinfo{person}{Sam Gross}, \bibinfo{person}{Francisco Massa}, \bibinfo{person}{Adam Lerer}, \bibinfo{person}{James Bradbury}, \bibinfo{person}{Gregory Chanan}, \bibinfo{person}{Trevor Killeen}, \bibinfo{person}{Zeming Lin}, \bibinfo{person}{Natalia Gimelshein}, \bibinfo{person}{Luca Antiga}, {et~al\mbox{.}}} \bibinfo{year}{2019}\natexlab{}.
\newblock \showarticletitle{{PyTorch:} An imperative style, high-performance deep learning library}.
\newblock \bibinfo{journal}{\emph{Advances in Neural Information Processing Systems}}  \bibinfo{volume}{32} (\bibinfo{year}{2019}).
\newblock


\bibitem[Pathak et~al\mbox{.}(2022)]%
        {pathak2022fourcastnet}
\bibfield{author}{\bibinfo{person}{Jaideep Pathak}, \bibinfo{person}{Shashank Subramanian}, \bibinfo{person}{Peter Harrington}, \bibinfo{person}{Sanjeev Raja}, \bibinfo{person}{Ashesh Chattopadhyay}, \bibinfo{person}{Morteza Mardani}, \bibinfo{person}{Thorsten Kurth}, \bibinfo{person}{David Hall}, \bibinfo{person}{Zongyi Li}, \bibinfo{person}{Kamyar Azizzadenesheli}, {et~al\mbox{.}}} \bibinfo{year}{2022}\natexlab{}.
\newblock \showarticletitle{{FourCastNet}: A global data-driven high-resolution weather model using adaptive {F}ourier neural operators}.
\newblock \bibinfo{journal}{\emph{arXiv preprint arXiv:2202.11214}} (\bibinfo{year}{2022}).
\newblock


\bibitem[Petersen et~al\mbox{.}(2019)]%
        {petersen2019}
\bibfield{author}{\bibinfo{person}{Mark~R. Petersen}, \bibinfo{person}{Xylar~S. Asay-Davis}, \bibinfo{person}{Anne~S. Berres}, \bibinfo{person}{Qingshan Chen}, \bibinfo{person}{Nils Feige}, \bibinfo{person}{Matthew~J. Hoffman}, \bibinfo{person}{Douglas~W. Jacobsen}, \bibinfo{person}{Philip~W. Jones}, \bibinfo{person}{Mathew~E. Maltrud}, \bibinfo{person}{Stephen~F. Price}, \bibinfo{person}{Todd~D. Ringler}, \bibinfo{person}{Gregory~J. Streletz}, \bibinfo{person}{Adrian~K. Turner}, \bibinfo{person}{Luke~P. Van~Roekel}, \bibinfo{person}{Milena Veneziani}, \bibinfo{person}{Jonathan~D. Wolfe}, \bibinfo{person}{Phillip~J. Wolfram}, {and} \bibinfo{person}{Jonathan~L. Woodring}.} \bibinfo{year}{2019}\natexlab{}.
\newblock \showarticletitle{An Evaluation of the Ocean and Sea Ice Climate of {E3SM using MPAS} and Interannual {CORE-II} Forcing}.
\newblock \bibinfo{journal}{\emph{Journal of Advances in Modeling Earth Systems}} \bibinfo{volume}{11}, \bibinfo{number}{5} (\bibinfo{year}{2019}), \bibinfo{pages}{1438--1458}.
\newblock
\urldef\tempurl%
\url{https://doi.org/10.1029/2018MS001373}
\showDOI{\tempurl}
\showeprint{https://agupubs.onlinelibrary.wiley.com/doi/pdf/10.1029/2018MS001373}


\bibitem[Raissi et~al\mbox{.}({[n.\,d.]})]%
        {raissi_physics_2017}
\bibfield{author}{\bibinfo{person}{Maziar Raissi}, \bibinfo{person}{Paris Perdikaris}, {and} \bibinfo{person}{George~Em Karniadakis}.} \bibinfo{year}{[n.\,d.]}\natexlab{}.
\newblock \bibinfo{title}{Physics Informed Deep Learning {(Part I)}: Data-driven Solutions of Nonlinear Partial Differential Equations}.
\newblock
\newblock
\urldef\tempurl%
\url{https://doi.org/10.48550/arXiv.1711.10561}
\showDOI{\tempurl}
\showeprint[arxiv]{1711.10561 [cs, math, stat]}


\bibitem[Rashid et~al\mbox{.}(2022)]%
        {rashid2022learning}
\bibfield{author}{\bibinfo{person}{Meer~Mehran Rashid}, \bibinfo{person}{Tanu Pittie}, \bibinfo{person}{Souvik Chakraborty}, {and} \bibinfo{person}{NM~Anoop Krishnan}.} \bibinfo{year}{2022}\natexlab{}.
\newblock \showarticletitle{Learning the stress-strain fields in digital composites using Fourier neural operator}.
\newblock \bibinfo{journal}{\emph{Iscience}} \bibinfo{volume}{25}, \bibinfo{number}{11} (\bibinfo{year}{2022}).
\newblock


\bibitem[Redi(1982)]%
        {Redi1982}
\bibfield{author}{\bibinfo{person}{Martha~H. Redi}.} \bibinfo{year}{1982}\natexlab{}.
\newblock \showarticletitle{Oceanic Isopycnal Mixing by Coordinate Rotation}.
\newblock \bibinfo{journal}{\emph{Journal of Physical Oceanography}} \bibinfo{volume}{12}, \bibinfo{number}{10} (\bibinfo{year}{1982}), \bibinfo{pages}{1154--1158}.
\newblock
\urldef\tempurl%
\url{https://doi.org/10.1175/1520-0485(1982)012<1154:OIMBCR>2.0.CO;2}
\showDOI{\tempurl}


\bibitem[Ringler et~al\mbox{.}(2013)]%
        {RINGLER2013211}
\bibfield{author}{\bibinfo{person}{Todd Ringler}, \bibinfo{person}{Mark Petersen}, \bibinfo{person}{Robert~L. Higdon}, \bibinfo{person}{Doug Jacobsen}, \bibinfo{person}{Philip~W. Jones}, {and} \bibinfo{person}{Mathew Maltrud}.} \bibinfo{year}{2013}\natexlab{}.
\newblock \showarticletitle{A multi-resolution approach to global ocean modeling}.
\newblock \bibinfo{journal}{\emph{Ocean Modelling}}  \bibinfo{volume}{69} (\bibinfo{year}{2013}), \bibinfo{pages}{211--232}.
\newblock
\showISSN{1463-5003}
\urldef\tempurl%
\url{https://doi.org/10.1016/j.ocemod.2013.04.010}
\showDOI{\tempurl}


\bibitem[Ronneberger et~al\mbox{.}(2015)]%
        {ronneberger2015u}
\bibfield{author}{\bibinfo{person}{Olaf Ronneberger}, \bibinfo{person}{Philipp Fischer}, {and} \bibinfo{person}{Thomas Brox}.} \bibinfo{year}{2015}\natexlab{}.
\newblock \showarticletitle{U-net: Convolutional networks for biomedical image segmentation}. In \bibinfo{booktitle}{\emph{Medical Image Computing and Computer-Assisted Intervention--MICCAI 2015: 18th International Conference, Munich, Germany, October 5-9, 2015, Proceedings, Part III 18}}. Springer, \bibinfo{pages}{234--241}.
\newblock


\bibitem[Semtner(1995)]%
        {semtner1995modeling}
\bibfield{author}{\bibinfo{person}{Albert~J. Semtner}.} \bibinfo{year}{1995}\natexlab{}.
\newblock \showarticletitle{Modeling Ocean Circulation}.
\newblock \bibinfo{journal}{\emph{Science}} \bibinfo{volume}{269}, \bibinfo{number}{5229} (\bibinfo{year}{1995}), \bibinfo{pages}{1379--1385}.
\newblock
\urldef\tempurl%
\url{https://doi.org/10.1126/science.269.5229.1379}
\showDOI{\tempurl}


\bibitem[Sun et~al\mbox{.}(2023)]%
        {sun2023deepgraphonet}
\bibfield{author}{\bibinfo{person}{Yixuan Sun}, \bibinfo{person}{Christian Moya}, \bibinfo{person}{Guang Lin}, {and} \bibinfo{person}{Meng Yue}.} \bibinfo{year}{2023}\natexlab{}.
\newblock \showarticletitle{{DeepGraphONet}: A deep graph operator network to learn and zero-shot transfer the dynamic response of networked systems}.
\newblock \bibinfo{journal}{\emph{IEEE Systems Journal}} (\bibinfo{year}{2023}).
\newblock


\bibitem[Thiyagalingam et~al\mbox{.}(2022)]%
        {thiyagalingam2022scientific}
\bibfield{author}{\bibinfo{person}{Jeyan Thiyagalingam}, \bibinfo{person}{Mallikarjun Shankar}, \bibinfo{person}{Geoffrey Fox}, {and} \bibinfo{person}{Tony Hey}.} \bibinfo{year}{2022}\natexlab{}.
\newblock \showarticletitle{Scientific machine learning benchmarks}.
\newblock \bibinfo{journal}{\emph{Nature Reviews Physics}} \bibinfo{volume}{4}, \bibinfo{number}{6} (\bibinfo{year}{2022}), \bibinfo{pages}{413--420}.
\newblock


\bibitem[von Frese et~al\mbox{.}(1997)]%
        {von1997analysis}
\bibfield{author}{\bibinfo{person}{Ralph~RB von Frese}, \bibinfo{person}{Michael~B Jones}, \bibinfo{person}{Jeong~Woo Kim}, {and} \bibinfo{person}{Jeong-Hee Kim}.} \bibinfo{year}{1997}\natexlab{}.
\newblock \showarticletitle{Analysis of anomaly correlations}.
\newblock \bibinfo{journal}{\emph{Geophysics}} \bibinfo{volume}{62}, \bibinfo{number}{1} (\bibinfo{year}{1997}), \bibinfo{pages}{342--351}.
\newblock


\bibitem[Wolfram et~al\mbox{.}(2015)]%
        {Wolfram2015}
\bibfield{author}{\bibinfo{person}{Phillip~J. Wolfram}, \bibinfo{person}{Todd~D. Ringler}, \bibinfo{person}{Mathew~E. Maltrud}, \bibinfo{person}{Douglas~W. Jacobsen}, {and} \bibinfo{person}{Mark~R. Petersen}.} \bibinfo{year}{2015}\natexlab{}.
\newblock \showarticletitle{Diagnosing Isopycnal Diffusivity in an Eddying, Idealized Midlatitude Ocean Basin via {Lagrangian}, in Situ, Global, High-Performance Particle Tracking {(LIGHT)}}.
\newblock \bibinfo{journal}{\emph{Journal of Physical Oceanography}} \bibinfo{volume}{45}, \bibinfo{number}{8} (\bibinfo{year}{2015}), \bibinfo{pages}{2114--2133}.
\newblock
\urldef\tempurl%
\url{https://doi.org/10.1175/JPO-D-14-0260.1}
\showDOI{\tempurl}


\bibitem[Yan et~al\mbox{.}(2018)]%
        {yan2018underestimated}
\bibfield{author}{\bibinfo{person}{Xiaoqin Yan}, \bibinfo{person}{Rong Zhang}, {and} \bibinfo{person}{Thomas~R. Knutson}.} \bibinfo{year}{2018}\natexlab{}.
\newblock \showarticletitle{Underestimated {AMOC} Variability and Implications for {AMV} and Predictability in {CMIP} Models}.
\newblock \bibinfo{journal}{\emph{Geophysical Research Letters}} \bibinfo{volume}{45}, \bibinfo{number}{9} (\bibinfo{year}{2018}), \bibinfo{pages}{4319--4328}.
\newblock
\urldef\tempurl%
\url{https://doi.org/10.1029/2018GL077378}
\showDOI{\tempurl}


\end{thebibliography}


\end{document}